\documentclass[prl,twocolumn,showpacs,preprintnumbers,amsmath,amssymb]{revtex4}

\usepackage{graphicx}% Include figure files
\usepackage{dcolumn}% Align table columns on decimal point
\usepackage{bm}% bold math

\begin{document}

\title{Adiabatic Mach-Zehnder Interferometry on A Quantized Bose-Josephson Junction}

\author{Chaohong Lee}
 \altaffiliation{Electronic addresses: chl124@rsphysse.anu.edu.au; chleecn@gmail.com}

\affiliation{Nonlinear Physics Centre and ARC Centre of Excellence
for Quantum-Atom Optics, Research School of Physical Sciences and
Engineering, Australian National University, Canberra ACT 0200,
Australia}

\date{\today}

\begin{abstract}

We propose a scheme to achieve a Mach-Zehnder interferometry using
a quantized Bose-Josephson junction with negative charging energy.
The quantum adiabatic evolution through a dynamical bifurcation is
used to accomplish the beam splitting and recombination. The
negative charging energy ensures the existence of a path-entangled
state which enhances the phase measurement precision to the
Heisenberg limit. A feasible detection procedure is also
presented. The scheme should be realizable with current
technology.

\end{abstract}


\maketitle

Quantum interference, one of the most fundamental and challenging
principles in quantum mechanics, forms the basis of the high
precision measurement and the quantum information processing. A
basic device capable of performing high precision measurements is
an analogue to the optical Mach-Zehnder interferometer, in which,
an incoming wave is divided into two parts by a 50:50 beamsplitter
and then the two parts are recombined by another beamsplitter.
Conventional Mach-Zehnder interferometry utilizing single-particle
states can only reach the standard quantum limit or the shot noise
limit on the measurement precision \cite{QEM}. However, it has
been demonstrated that many-body quantum entanglement, such as the
photon polarization entanglement \cite{PhotonPE} and the
trapped-ion internal entanglement \cite{TIonE}, can enhance the
measurement precision to the so-called Heisenberg limit \cite{QEM}
posed by the Heisenberg uncertainty principle. The application of
spin entanglement in interferometer has also been discussed
\cite{SSS}.

Given the well-developed techniques in preparing and manipulating
atomic Bose-Einstein condensates, the interferometric schemes
based upon ultracold atoms stimulate great interests. Using
spatially separated condensates \cite{CI-separated-BEC},
condensates trapped within double-well potentials
\cite{CI-doublewell-BEC} and classical Josephson arrays of
tunnelling coupled condensates \cite{CI-BEC-lattices}, the atomic
coherence and interference has been demonstrated . All these
experiments have utilized the macroscopic quantum coherence which
is well described within the mean-field theory. The many-body
nature becomes significant for strongly correlated atoms and some
many-body quantum effects including quantum squeezing
\cite{squeezing}, quantum entanglement \cite{entanglement} and
quantum phase transition \cite{QPT} have been explored. These
effects lie beyond the reaches of the mean-field theory.

On the other hand, due to the s-wave scattering dominates the
ultracold collisions, the nonlinear Kerr effect is intrinsic to
the atomic condensates. It has been shown that the nonlinear
interaction brings a number of novel phenomena
\cite{Nature-Nonlinear-atom-optics} including wave mixing
\cite{mixing}, soliton \cite{soliton}, dynamical bifurcation
\cite{bifurcation} and chaos \cite{chaos}. In this Letter we will
demonstrate how the combination of nonlinear and many-body quantum
effects can be used to realize a Heisenberg-limited Mach-Zehnder
interferometry with Bose condensed atoms.

In this Letter, in the frame of fully quantized theory, we propose
and analyze a practical scheme of a Mach-Zehnder interferometry
with a Bose-Josephson junction. The beamsplitters are realized by
the quantum adiabatic processes through the dynamical bifurcation.
To ensure the existence of a path-entangled state, which enhances
the phase measurement precision to the Heisenberg limit, the
charging energy is chosen to be negative values. We also discuss a
feasible procedure for detection and the experimental realization.
The proposed interferometry scheme can operate for large particle
numbers ($\sim 10^{3}$). Whereas the schemes of photons
\cite{PhotonPE} and trapped ions \cite{TIonE} can only reach the
order of $10$ particles. Reduced influence of the environment and
a simple detection procedure are advantages of this scheme.

We consider an ensemble of $N$ Bose condensed atoms confined in a
double-well potential [or $N$ two-level Bose condensed atoms
confined in a single-well potential with laser (or radio
frequency) coupling between the two involved hyperfine levels].
Under the condition of tight-binding, the system obeys a two-mode
Hamiltonian,
$$
H=\frac{\delta}{2} (n_{2}-n_{1}) -
\frac{T}{2} (a^{+}_{2}a_{1}+a^{+}_{1}a_{2}) +
\frac{E_{C}}{8}(n_{2}-n_{1})^{2}.
$$
Here, $a^{+}_{j}$, $a_{j}$ and $n_{j}=a^{+}_{j}a_{j}$ $(j=1, 2)$
denote the creation, annihilation and particle number operators
for the $j$-th mode, respectively. This Hamiltonian describes a
quantized Bose-Josephson junction with an imbalance $\delta$, an
inter-mode coupling $T$ and a charging energy $E_{C}$. The values
of $\delta$, $T$ and $E_{C}$ are controlled by the potential
asymmetry (or the internal energy difference), the tunnelling
strength between two wells (or the Rabi frequency for coupling
fields), and the s-wave scattering lengths, respectively
\cite{squeezing,entanglement,bifurcation,chaos,model}. Regarding
all atoms as spin-$1/2$ particles, one can define the angular
momentum operators as $J_{x}=(a^{+}_{2}a_{1}+a_{2}a^{+}_{1})/2$,
$J_{y}=i(a^{+}_{2}a_{1}-a_{2}a^{+}_{1})/2$ and
$J_{z}=(a^{+}_{2}a_{2}-a^{+}_{1}a_{1})/2$. Thus, $H=\delta J_{z} -
T J_{x} + E_{C} J_{z}^{2} /2$ and an arbitrary state can be
expanded as a superposition of different states $\left | J=N/2,
J_{z}=M \right \rangle$ with $M=-N/2, -N/2+1, \cdots, +N/2$.

The ground states of the quantized Bose-Josephson junction
sensitively depend on the parameters. For a symmetric junction
($\delta=0$), in the strong coupling limit ($T/|E_{C}| \gg 1$),
the ground state is a SU(2) spin coherent state $\exp(i \phi
J_{z}) \exp(i \theta J_{y})|N/2,+N/2\rangle$ with $\phi =0$ and
$\theta =\pi/2$. In the weak coupling limit ($T/|E_{C}| \ll 1$),
the ground state relies on $E_{C}$. If $E_{C}>0$, the ground state
approaches to $(|N/2,-1/2\rangle + |N/2,+1/2\rangle)/\sqrt{2}$ for
odd $N$ or $|N/2,0\rangle$ for even $N$, when $T\rightarrow 0$. If
$E_{C}<0$, the ground state $|0\rangle$ and the first excited
state $|1\rangle$ become degenerate when $T\rightarrow 0$, as
shown in Fig. 1 (a). The critical value between nondegeneracy and
degeneracy corresponds to a classical Hopf bifurcation from
single-stability to bistability \cite{bifurcation}. With $T=0$,
these two states are the lowest spin state $|N/2,-N/2\rangle$ and
the highest spin state $|N/2,+N/2\rangle$, respectively. However,
the degeneracy between $|0\rangle$ and $|1\rangle$ will be
destroyed by the appearance of a nonzero $\delta$. In Fig. 1, we
show the energy spectra and the ground states for a quantized
Bose-Josephson junction with $\delta=0$, $E_{C}=-2.0$ and $N=20$.
%%%%%%%%%%%%%%%%%%%%%%%%%%%%%%%%%%%%%%%%%%%%%%%%%%%%%%%%%%%%%%%%
\begin{figure}[ht]
\rotatebox{0}{\resizebox *{\columnwidth}{5.5cm} {\includegraphics
{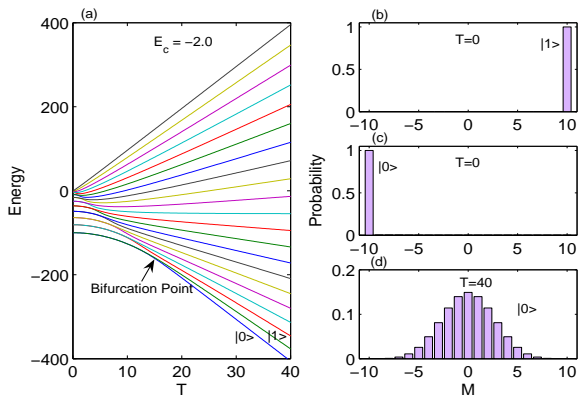}}} \caption{The energy spectra and the ground states for
a symmetric quantized Bose-Josephson junction with $E_{C}=-2.0$
and $N=20$. (a) The energy spectra for different $T$. (b) and (c)
The degenerated first excited state $|1\rangle$ and the ground
state $|0\rangle$ for $T=0$. (d) The ground state for $T=40$.}
\end{figure}
%%%%%%%%%%%%%%%%%%%%%%%%%%%%%%%%%%%%%%%%%%%%%%%%%%%%%%%%%%%%%%%%%

For the system of a larger $N$, similar transition from
nondegeneracy and degeneracy could be induced by the bifurcation.
Its energy spectrum can be analyzed with the Bethe ansatz
\cite{BA}. Its numerical simulation is difficult to perform, while
some approximate methods including the iteration diagonalization
and the density matrix renormalization group methods have been
suggested, such an analysis is beyond the scope of our paper. We
stress that the physics of the analyzed situation is the source
for the systems of small ($\sim 10$) and large ($\geq 10^{3}$) $N$
and the used exact diagonalization is an adequate tool for our
proof-of-the-principle calculations.

By using the ground state and the first excited state as two paths
of an interferometer and accomplishing the beam splitting and
recombination via adiabatic passage through dynamical bifurcation,
it is possible to realize a Mach-Zehnder interferometry on a
quantized Bose-Josephson junction with finite negative $E_{C}$.

The first beamsplitter can be achieved by preparing the ground
state in the strong coupling limit and then slowly decreasing $T$
to zero, due to the appearance of a dynamical bifurcation. The
negative $E_{C}$ ensures that one will get a path entangled state
$(|N/2,-N/2\rangle + |N/2,+N/2\rangle)/\sqrt{2}$ in the weak
coupling limit. So that the first beamsplitter also provides a
route to producing a kind of entangled states. Even in the
appearance of the Landau-Zener tunnelling induced by the
imbalance, if $|\delta|<\delta _{C}$, one can still get a path
entangled state $(|N/2,-N/2\rangle + e ^{i \varphi}
|N/2,+N/2\rangle)/\sqrt{2}$ with a desired high fidelity. The
phase difference $\varphi$ mainly comes from the imbalance and the
critical value $\delta _{C}$ depends on the parameters and the
desired fidelity. Utilizing the Landau-Zener tunnelling as
coherent beamsplitters, the Mach-Zehnder interferometry has been
demonstrated in a superconducting flux qubit
\cite{superconductor-MZ}. Similarly, starting from
$|N/2,-N/2\rangle$ or $|N/2,+N/2\rangle$ at $T=0$, and slowly
increasing $T$ to $T \gg |E_{C}|$, beam splitting is induced by
the bifurcation. In the strong coupling limit, one will obtain an
equal probability superposition state of $|0\rangle$ and
$|1\rangle$. In all these adiabatic processes, the evolving states
perfectly keep in the sub Hilbert space expanded by the ground
state and the first excited state.

For a quantized Bose-Josephson junction with $E_{C}=-2.0$, $N=20$
and $T=40-t$ (where $t$ is the evolution time), in Fig. 2, we show
the initial state, the destination states and the fidelities
$F_{0}=\left | \langle 0 | \Psi (t) \rangle \right |^{2}$ and
$F_{1}=\left | \langle 1 | \Psi (t) \rangle \right |^{2}$ of the
evolving state $\Psi (t)$ in a beam splitting process from $T=40$
to $0$. Here, $F_{0}$ and $F_{1}$ denote the populations of the
evolving states $\Psi (t)$ occupying the states $| 0 \rangle$ and
$| 1 \rangle$, respectively. The result shows the total fidelity
$F_{0} + F_{1}$ almost keeps unchanged, this indicates that  the
environment effect is dramatically suppressed in the involved
adiabatic processes. Controlling the imbalance $|\delta |<0.015$,
one can get the path entangled states
$|\Phi(\varphi)\rangle=(|N/2,-N/2\rangle + e ^{i \varphi}
|N/2,+N/2\rangle)/\sqrt{2}$ with high fidelities larger than 0.98,
see Fig. 2 (d).
%%%%%%%%%%%%%%%%%%%%%%%%%%%%%%%%%%%%%%%%%%%%%%%%%%%%%%%%%%%%%%%%
\begin{figure}[ht]
\rotatebox{0}{\resizebox *{\columnwidth}{5.5cm} {\includegraphics
{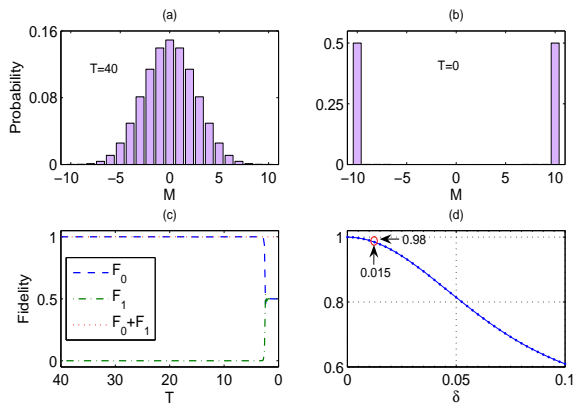}}} \caption{The beam splitting process from $T=40$ to
$T=0$ in a symmetric (cases a, b and c) or weakly asymmetric (case
d) quantized Bose-Josephson junction with $E_{C}=-2.0$ and $N=20$.
(a) The initial state at $T=40$ is the ground state. (b) The
destination state at $T=0$ is a path entangled state
$(|10,-10\rangle + |10,+10\rangle)/\sqrt{2}$. (c) The fidelities
of the evolving states $\Psi (T)$ versus the coupling $T$. (d) The
maximum fidelity $F^{Max}_{\Phi}=Max (\left | \langle
\Phi(\varphi) | \Psi (T=0, \delta) \rangle \right |^{2})$ of the
destination state $\Psi (T=0, \delta)$ to the path entangled state
$|\Phi(\varphi)\rangle$ versus the imbalance $\delta$.}
\end{figure}
%%%%%%%%%%%%%%%%%%%%%%%%%%%%%%%%%%%%%%%%%%%%%%%%%%%%%%%%%%%%%%%%%

Inducing an unknown phase shift $\phi$ between two paths with a
mode-dependent force, the path entangled state prepared by the
first beamsplitter evolves into a $\phi$-shifted path entangled
state $|\Phi(\phi)\rangle=(|N/2,-N/2\rangle + e ^{i \phi}
|N/2,+N/2\rangle)/\sqrt{2}$. To extract the information on the
phase shift, one has to recombine $|\Phi(\phi)\rangle$ by the
second beamsplitter and monitor the populations in the two output
paths. For a symmetric or a weakly asymmetric quantized
Bose-Josephson junction with a negative $E_{C}$, the beam
recombination can be achieved by a dynamical bifurcation or a
Landau-Zener tunnelling in the process of slowly increasing $T$
from $0$ to $T \gg |E_{C}|$. Finally, in the strong coupling
limit, the populations in the ground state and the first excited
state will show interference behavior with the outcome determined
by the phase shift $\phi$. In an ideal case, the fidelities of the
final state to the ground and the first excited states can be
exactly expressed as $F_{0}=\left | \langle 0 | \Psi (T) \rangle
\right |^{2}_{T \gg |E_{C}|}= \cos^{2}(\phi/2)$ and $F_{1}=\left |
\langle 1 | \Psi (T) \rangle \right |^{2}_{T \gg |E_{C}|}=
\sin^{2}(\phi/2)$, respectively. This means that all particles
will occupy the ground state if $\phi=2k \pi$ (where $k$ is an
integer) or will stay in the first excited state if $\phi=(2k+1)
\pi$. By slowly increasing $T$ from $0$ to $40$, we simulate the
beam recombination process in a symmetric quantized Bose-Josephson
junction with $N=20$ and $E_{C}=-2.0$ from the initial path
entangled sates $|\Phi(\phi)\rangle=(|10,-10\rangle + e ^{i \phi}
|10,+10\rangle)/\sqrt{2}$, see Fig. 3. The fidelities of the
destination state $\Psi (T=40)$ to the ground state $F_{0}$ and
the first excited state $F_{1}$ show perfect behaviors of the
Mach-Zehnder interference. Utilizing the path entangled states in
our scheme, the phase measurement precision reaches the Heisenberg
limit \cite{QEM}. This is in contrast to the conventional schemes
using untangled single-particle states which can only reach the
standard quantum limit or the shot noise limit \cite{QEM}.
%%%%%%%%%%%%%%%%%%%%%%%%%%%%%%%%%%%%%%%%%%%%%%%%%%%%%%%%%%%%%%%%
\begin{figure}[ht]
\rotatebox{0}{\resizebox *{\columnwidth}{5.3cm} {\includegraphics
{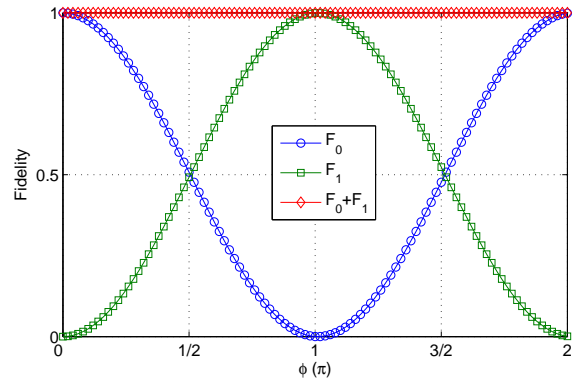}}} \caption{Behaviors of Mach-Zehnder interference
in the output states of a symmetric quantized Bose-Josephson
junction with $N=20$ and $E_{C}=-2.0$ with initial states
$|\Phi(\phi)\rangle$ through a dynamical bifurcation to the strong
coupling limit ($T=40$).}
\end{figure}
%%%%%%%%%%%%%%%%%%%%%%%%%%%%%%%%%%%%%%%%%%%%%%%%%%%%%%%%%%%%%%%%%

Due to the absence of an effective method to distinguish between
$| 0 \rangle$ and $| 1 \rangle$ in the strong coupling limit, it
is not easy to know the populations in these two states by
directly detecting the output state from the previous procedure.
Fortunately, utilizing the similarity of $| 0 \rangle$ and $| 1
\rangle$ for symmetric and asymmetric quantized Bose-Josephson
junctions in the strong coupling limit and the nondegeneracy of
these two states for an asymmetric junction, one can distinguish
these two states by suddenly applying a proper imbalance and then
slowly decreasing $T$ from $T \gg |E_{C}|$ to zero (or close to
zero). To determine the population of the states in the strong
coupling limit, one has to avoid the dynamical bifurcation and the
Landau-Zener tunnelling, that is to say, one has to maintain the
populations in both $| 0 \rangle$ and $| 1 \rangle$ unchanged. To
avoid the dynamical bifurcation, the applied imbalance must
satisfy the condition $|\delta|<|E_{C}|/2$; and to make the
Landau-Zener tunnelling absent, it is the best to choose the
imbalance $|\delta|=|E_{C}|/4$. Under these conditions, when $T$
approaches zero, the ground state and the first excited state
become the lowest spin state $|N/2,-N/2\rangle$ and the highest
spin state $|N/2,+N/2\rangle$, respectively. These two states with
the largest $|J_{z}|$ correspond to all particles completely
localized in either of the two modes, which can be easily
detected. For an asymmetric quantized Bose-Josephson junction with
$N=20$, $E_{C}=-2.0$, $\delta=0.5$, and $T>35$, the fidelities of
the ground state and the first excited state to the symmetric
counterparts are very close to one, i.e., $F_{00^{\prime}}=\left |
_{\delta=0}\langle 0^{\prime}|0 \rangle _{\delta=0.5}\right
|^{2}_{T>35}\simeq 1$ and $F_{11^{\prime}}=\left |
_{\delta=0}\langle 1^{\prime}|1 \rangle _{\delta=0.5}\right
|^{2}_{T>35}\simeq 1$, as seen in Fig. 4 (b). At $T=0$, the ground
state and the first excited state are the lowest spin state
$|10,-10\rangle$ [see Fig. 4 (c)] and the highest spin state
$|10,+10\rangle$ [see Fig. 4 (d)], respectively. Taking the output
states from the second beam splitter, suddenly applying an
imbalance $\delta=0.5$ and then slowly varying $T$ from 40 to 0,
our numerical simulation shows that the populations in the ground
state and the first excited states remain unchanged due to the
nondegeneracy and sufficiently large energy level distances.
%%%%%%%%%%%%%%%%%%%%%%%%%%%%%%%%%%%%%%%%%%%%%%%%%%%%%%%%%%%%%%%%
\begin{figure}[ht]
\rotatebox{0}{\resizebox *{\columnwidth}{5.5cm} {\includegraphics
{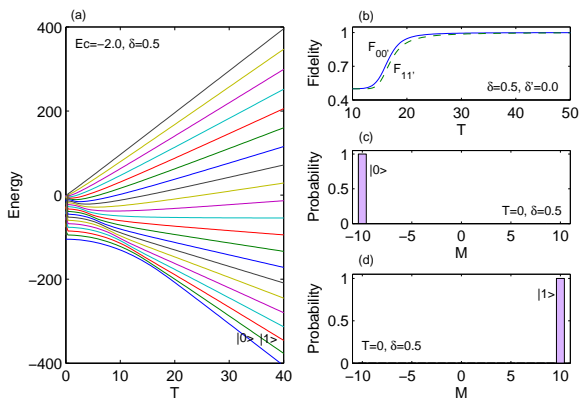}}} \caption{The energy spectrum, the states
$|0\rangle$ and $|1\rangle$ and their fidelities to the symmetric
counterparts for an asymmetric quantized Bose-Josephson junction
with $N=20$, $E_{C}=-2.0$ and $\delta=0.5$. (a) The energy
spectrum. (b) The fidelities $F_{00^{\prime}}$ and
$F_{11^{\prime}}$ for different $T$. (c) The ground state
$|0\rangle$ for $T=0$. (d) The first excited state $|1\rangle$ for
$T=0$.}
\end{figure}
%%%%%%%%%%%%%%%%%%%%%%%%%%%%%%%%%%%%%%%%%%%%%%%%%%%%%%%%%%%%%%%%%

There are two alternative approaches to the experimental
realization. One possibility is to trap the condensed atoms in a
double-well potential formed by external fields
\cite{CI-doublewell-BEC,model}, the other is to use a harmonically
trapped two-component condensate with internal coupling between
the two involved hyperfine levels \cite{model,TCBEC}. In both
cases, the negative charging energy can be obtained by controlling
the s-wave scattering lengths $a_{s}$ with Feshbach resonance. For
the condensates in double-well potentials, the imbalance and the
coupling strength can be precisely adjusted with the techniques
developed for matter-wave interference on an atom chip
\cite{CI-doublewell-BEC}. Assuming the experimental setup used to
observe self-trapping [Phys. Rev. Lett. \textbf{95}, 010402
(2005)] and changing $a_{s}$ from positive to negative, the
parameter $\lambda=N|E_{C}|/(2T) \sim 15$ ensures the appearance
of the path-entangled state. For the coupled two-component
condensates \cite{TCBEC}, the imbalance is determined by the
energy difference between the two involved hyperfine levels, and
can be varied by applying external magnetic fields. The coupling
strength is determined by the Rabi frequency of the coupling
fields, and can be controlled by changing the field intensity.
Assuming the experimental setup in [Phys. Rev. Lett. \textbf{81},
1539 (1998)], for $N=10^{3}$, $\lambda=N|E_{C}|/(2T) = 10$
requires the Rabi frequency $\Omega = T/\hbar \sim 2\pi$ Hz and
the exact path entangled state is reached at $\Omega =0$ Hz. To
count the atom numbers in a particular mode, one can use
fluorescence imaging for small total numbers of atoms or use
spatial imaging for large total numbers of atoms ($\sim 10^{3}$ or
larger).

To conclude, we have discussed a simple and robust scheme to
achieve a Heisenberg-limited Mach-Zehnder interferometry with
macroscopic many-body quantum states in a quantized Bose-Josephson
junction with negative charging energy, which can be realized with
the present level of expertise in manipulating ultracold atoms.
The interferometry not only is of fundamental physical interests,
but also offers possible technological applications in
high-precision measurements \cite{QEM,atom-coherence}, in which
the phase measurement precision reaches the Heisenberg limit posed
by the uncertainty principle. On the other hand, given the
routinely prepared condensates of $10^{3}$ or larger number of
atoms, the first beamsplitter in our interferometer also provides
an opportunity to produce path entangled states of very large
number of particles. The presented Mach-Zehnder interferometry
with path entangled states of large number of particles has an
obvious advantage over the schemes using entangled states of
photons \cite{PhotonPE} or trapped ions \cite{TIonE}, which can
only operate with the order of 10 particles. All involved
processes, except for the beam splitting and recombination,
satisfy the adiabatic condition, and all involved quantum states
are eigenstates of the system. Moreover, the sub Hilbert space of
the evolving states is closed even in procedures of beam splitting
and recombination, which means that our scheme ultimately
suppresses the experimental errors causing from the environment.
Additionally, unlike the interferometry on linear optics, which
needs one polarization detector per photon \cite{PhotonPE}, our
interferometry with a quantized Bose-Josephson junction needs only
two detectors for any number of particles.

The author acknowledges discussions with Yuri Kivshar and Elena
Ostrovskaya. This work is supported by the Australian Research
Council (ARC).

\end{document}